# Reversible spin texture in ferroelectric HfO$_2$

L. L. Tao,* Tula R. Paudel, Alexey A. Kovalev, and Evgeny Y. Tsymbal†

*Department of Physics and Astronomy & Nebraska Center for Materials and Nanoscience*
*University of Nebraska, Lincoln, Nebraska 68588, USA*

Spin-orbit coupling effects occurring in non-centrosymmetric materials are known to be responsible for non-trivial spin configurations and a number of emergent physical phenomena. Ferroelectric materials may be especially interesting in this regard due to reversible spontaneous polarization making possible for a non-volatile electrical control of the spin degrees of freedom. Here, we explore a technologically relevant oxide material, HfO$_2$, which has been shown to exhibit robust ferroelectricity in a non-centrosymmetric orthorhombic phase. Using theoretical modelling based on density-functional theory, we investigate the spin-dependent electronic structure of the ferroelectric HfO$_2$ and demonstrate the appearance of chiral spin textures driven by spin-orbit coupling. We analyze these spin configurations in terms of the Rashba and Dresselhaus effects within the $\mathbf{k}\cdot\mathbf{p}$ Hamiltonian model and find that the Rashba-type spin texture dominates around the valence band maximum, while the Dresselhaus-type spin texture prevails around the conduction band minimum. The latter is characterized by a very large Dresselhaus constant $\alpha_D = 0.578$ eV Å, which allows using this material as a tunnel barrier to produce tunneling anomalous and spin Hall effects that are reversible by ferroelectric polarization.

## I. INTRODUCTION

Crystalline materials lacking space inversion symmetry exhibit electronic energy bands that are split by spin-orbit coupling (SOC). This is due to a non-vanishing gradient of the electrostatic potential coupled to the electron spin through the intra-atomic SOC. As a result, in non-centrosymmetric crystals the SOC is odd in the electron's wave vector (**k**), as was first demonstrated by Dresselhaus [1] and Rashba. [2] The spin-momentum coupling lifts Kramers' spin degeneracy and leads to a complex **k**-dependent spin texture of the electronic bands. The Rashba and Dresselhaus effects have recently aroused significant interest in conjunction to thin-film heterostructures where a number of emergent physical phenomena are triggered by these SOC effects.[3] A particular interest is driven due to a unique possibility to manipulate the spin degrees by an external electric field, [4,5] which is of great importance for spintronics – a field of research promising revolutionize future electronics. [6]

The Rashba effect has been observed on surfaces and interfaces where space inversion symmetry is violated due to the structural confinement. For example, surfaces of heavy metals, such as Au (111) [7] and Bi (111), [8] surfaces of oxides, such as SrTiO$_3$ (001) [9] and KTaO$_3$ (001),[10] two-dimensional materials, [11-13] and heterostructure interfaces, such as InGaAs/InAlAs [14] and LaAlO$_3$/SrTiO$_3$, [15] were demonstrated to exhibit the Rashba splitting. The giant Rashba effect has also been predicted and observed in bulk materials, such as BiTeI,[16,17] and GeTe. [18,19] The Dresselhaus effect was originally proposed for bulk zinc-blende and wurtzite semiconductors, where the spin splitting was predicted to be proportional to $k^3$.[1] The spin-momentum coupling linear in $k$ can also be realized in non-centrosymmetric structures giving rise to the linear Dresselhaus SOC. [20] For example, the linear Dresselhaus term was found to be sizable for indirect-gap zinc-blende semiconductors, such as AlAs and GaP. [21]

The linear SOC can be written in terms of an effective **k**-dependent field $\mathbf{\Omega}(\mathbf{k})$ affecting the spin $\mathbf{\sigma}$: [22]

$$H_{SO} = \mathbf{\Omega}(\mathbf{k})\cdot\mathbf{\sigma}, \qquad (1)$$

where $\mathbf{\Omega}(\mathbf{k})$ is linear in **k** and thus $H_{SO}$ preserves the time-reversal symmetry. The specific form of $\mathbf{\Omega}(\mathbf{k})$ depends on the space symmetry of the system. For example, in case of the $C_{2v}$ point group, the Dresselhaus and Rashba SOC fields can be written as $\mathbf{\Omega}_D(\mathbf{k}) = \lambda_D(k_y, k_x, 0)$ and $\mathbf{\Omega}_R(\mathbf{k}) = \lambda_R(k_y, -k_x, 0)$, respectively.

Among non-centrosymmetric materials exhibiting Rashba and Dresselhaus effects are ferroelectrics, which are characterized by spontaneous polarization switchable by an applied electric field. It was proposed that in such materials a full reversal of the spin texture can occur in response to the reversal of ferroelectric polarization. [23] Such functionality is interesting in view of potential technological applications employing, for example, tunneling anomalous and spin Hall effects,[24,25] controlled by ferroelectric polarization.

The original proposal explored properties of ferroelectric semiconductor GeTe.[18,23] Following this theoretical prediction, a number of other materials were considered as possible candidates for electrically switchable spin texture. Among them are metallo-organic halide perovskites, such as (FA)SnI$_3$,[26-28] hexagonal semiconductors, such as LiZnSb, [29] strained KTaO$_3$, [30] and BiAlO$_3$. [31] Coexistence of the Rashba and Dresselhaus SOC effects was predicted for ferroelectric (FA)SnI$_3$ [26] and BiAlO$_3$.[31]

Despite these advances, several challenges impede experimental studies and practical applications of the proposed materials. In particular, GeTe has a relatively small band gap (~0.5 eV), which leads to large conductivity hindering the ferroelectric switching process. Halide perovskites, on the other hand, suffer from limited structural stability and could hardly be integrated in the modern semiconductor technologies. Also ferroelectricity of these compounds is questionable. The proposed oxide materials require strain ($KTaO_3$) or high temperature/pressure ($BiAlO_3$) to be synthetized. It would be desirable to find a robust ferroelectric material, which suffices the requirements of practical application.

Hafnia ($HfO_2$) is a promising candidate for this purpose. This material is considered as a favorable gate dielectric in the metal–oxide–semiconductor field-effect transistor (MOSFET). This is due to its high dielectric constant (~25), a large band gap (~5.7 eV) that suppresses the leakage current, and good compatibility with Si. Recently, it was found that thin films of doped hafnia exhibit pronounced ferroelectric properties,[32-34] which makes this material promising also for applications in the ferroelectric field effect transistors and memories,[35] as well as ferroelectric tunnel junctions (FTJs).[36-40] The origin of the ferroelectric behavior was attributed to the formation of a non-centrosymmetric orthorhombic phase.[32,33] Based on a first-principles search algorithm, two possible ferroelectric phases were identified in $HfO_2$, namely orthorhombic polar phases with space group symmetries of $Pca2_1$ and $Pmn2_1$.[41] The direct experimental evidence of the ferroelectric $Pca2_1$ phase was recently provided by the scanning transmission electron microscopy.[42]

The ferroelectric phase of $HfO_2$ is interesting due to broken inversion symmetry which allows for the Rashba or Dresselhaus effects. Owing to Hf, which is a heavy 5d element, a sizable SOC is expected in this material, raising a natural question about magnitude of these effects. In this paper, we focus on the orthorhombic $Pca2_1$ structural phase of $HfO_2$, which was proposed by the theory[41] and identified in the experiment.[42] Using density-functional theory (DFT) calculations, we predict the formation of chiral spin textures driven by the Rashba and Dresselhaus effects. The spin textures are fully reversible with ferroelectric polarization, which makes this material promising for novel spintronic applications.

The rest of the paper is organized as follows. In section II, we describe details of the computational methods. Section III is devoted to the structural properties of ferroelectric $HfO_2$. Section IV is focused on the electronic structure and analysis of band symmetry. The spin textures are analyzed in terms of the Rashba and Dresselhaus effects within DFT calculations in section V and a model Hamiltonian approach in section VI. In section VII, we discuss some implications of our results, which are summarized in section VIII.

## II. COMPUTATIONAL METHODS

We employ DFT calculations utilizing the plane-wave ultrasoft pseudopotential method[43] implemented in Quantum-ESPRESSO.[44] The exchange-correlation functional is treated in the generalized gradient approximation (GGA).[45] Self-consistent computations are performed using an energy cutoff of 680 eV for the plane wave expansion and 10×10×10 Monkhorst-Pack grid for $k$-point sampling. A 16×16×16 k-point mesh is used for the calculation of the density of states. Atomic relaxations are performed in the absence of SOC until the Hellmann-Feynman forces on each atom became less than 2.6 meV/Å. The ferroelectric polarization is computed using the Berry phase method.[46] The expectation values of the spin operators $s_\alpha$ ($\alpha = x, y, z$) are found from

$$s_\alpha = \tfrac{1}{2}\langle \psi_k | \sigma_\alpha | \psi_k \rangle ,  \qquad (2)$$

where $\sigma_\alpha$ are the Pauli spin matrices and $\psi_k$ is the spinor eigenfunction, which is obtained from the non-collinear spin calculations.

## III. ATOMIC STRUCTURE

We consider a bulk ferroelectric $HfO_2$ crystal which belongs to the orthorhombic phase of space group $Pca2_1$. This space group is non-symmorphic, i.e. possesses point-symmetry operations combined with non-primitive translations.[47] The $Pca2_1$ group contains four symmetry operations: the identity operation ($E$); two-fold screw rotation $S_z$ which consists of $\pi/2$ rotation around the $z$ axis followed by $c/2$ translation along the $z$ axis:

$$S_z : (x, y, z) \rightarrow (-x, -y, z + \tfrac{1}{2}c) ;  \qquad (3)$$

glide reflection $M_1$ which consists of reflection about the $y = 0$ plane followed by $\tfrac{1}{2}a$ translation along the $x$-axis:

$$M_1 : (x, y, z) \rightarrow (x + \tfrac{1}{2}a, -y, z) ;  \qquad (4)$$

and glide reflection $M_2$ which consists of reflection about the $x = \tfrac{1}{4}a$ plane followed by $\tfrac{1}{2}c$ translation along the $z$-axis:

$$M_2 : (x, y, z) \rightarrow (-x + \tfrac{1}{2}a, y, z + \tfrac{1}{2}c) .  \qquad (5)$$

Here $a$, $b$, and $c$ are the lattice constants. Fig. 1a shows the atomic structure of the orthorhombic $HfO_2$.

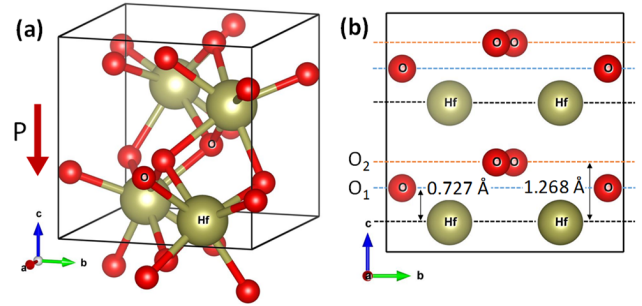

**FIG. 1.** (a) Crystal structure of bulk $HfO_2$ in the $Pca2_1$ orthorhombic phase. The polarization $P$ is along the $[00\bar{1}]$ direction as indicated by the red arrow. (b) Projection of the crystal structure to the (100) plane. Distances along the $c$ direction between Hf and O atomic planes are shown.

The $Pca2_1$ phase of HfO$_2$ is characterized by ferroelectric polarization parallel to the *c* axis as follows from the $C_{2v}$ point group symmetry corresponding to this space group. The $C_{2v}$ point group contains mirror *xz*- and *yz*-planes, which imply zero net polarization along the [100] or [010] directions. On the other hand, reflection about the *xy*-plane does not belong to this point group. The polar displacements between Hf and O ions along the *c*-axis yield a finite polarization pointing in the [00$\bar{1}$] direction, as is evident from Fig. 1(b), showing projection of the crystal structure to the (100) plane. There are two topologically equivalent variants of the space group $Pca2_1$ with opposite polarization (pointing in the [001] or [00$\bar{1}$] directions) indicative to the ferroelectric nature of HfO$_2$ in this crystallographic phase.

**Table 1.** Relaxed lattice constants and atomic positions for bulk HfO$_2$.

| Space group (No) | Lattice constants (Å) | | |
|---|---|---|---|
| $Pca2_1$ (29) | $a = 5.234$, $b = 5.010$, $c = 5.043$ | | |
| Atom | Wyckoff | *x* | *y* | *z* |
| Hf | 4*a* | 0.9668 | 0.7337 | 0.1231 |
| O1 | 4*a* | 0.1350 | 0.0666 | 0.2672 |
| O2 | 4*a* | 0.7288 | 0.4633 | 0.8744 |

Table 1 summarizes the calculated structural parameters (in the Wyckoff notation) for bulk HfO$_2$ in the orthorhombic $Pca2_1$ phase, which are in good agreement with the previous results.[41,48] The calculated polarization of 73 μC/cm$^2$ is also in line with the previous reported value of 52 μC/cm$^2$. [41]

## IV. ELECTRONIC STRUCTURE

Now we investigate the electronic structure of ferroelectric HfO$_2$. First, we perform non spin-polarized calculations, i.e. without including spin and SOC. Fig. 2a shows the calculated band dispersions along the selected *k* lines in the first Brillouin zone (shown in the inset). It is evident that HfO$_2$ is an indirect-band-gap insulator with the valence band maximum (VBM) located at the Γ point and the conduction band minimum (CBM) located near the high symmetry T point (0, π/*b*, π/*c*). The calculated band gap is about 4.6 eV. This is less than the reported experimental value of about 5.7 eV, due to the well-known deficiency of DFT to describe excited states. Fig. 2b shows the partial density of states (DOS) projected onto O-2*p* orbitals and Hf-4*d* orbitals. We see that the valence bands are mainly composed of the O-2*p* orbitals, whereas the conduction bands are mainly formed from the Hf-4*d* orbitals.

Our calculations find that the electronic bands are double degenerate along the symmetry lines Z – U – R – T – Z lying at the Brillouin zone boundary plane $k_z = \pi/c$ (highlighted in green in inset of Fig. 2). This double degeneracy is not related to the *d*-orbital character of the conduction band. For example, the states around the T point have $e_g$ symmetry and represent two singlets of the $d_{z^2}$ and $d_{x^2-y^2}$ character (denoted by $T_1$ and $T_1'$, respectively, in Fig. 2). The double degeneracy of the bands in the $k_z = \pi/c$ plane is protected by the non-symmorphic symmetry of the $Pca2_1$ space group of the HfO$_2$ orthorhombic phase. This can be understood as follows. [49,50]

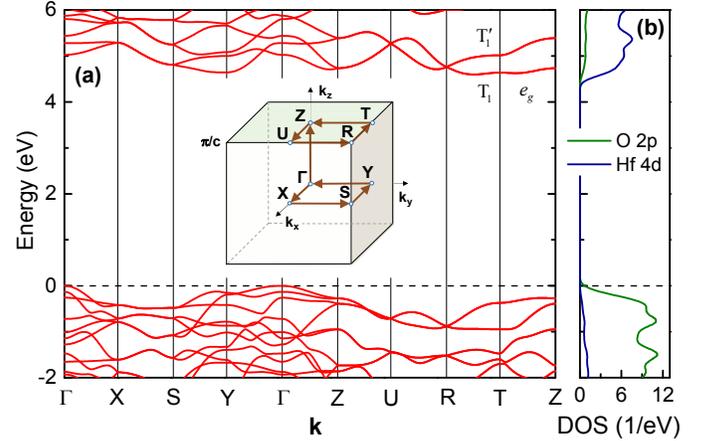

**FIG. 2.** Electronic band structure of HfO$_2$ in absence of SOC. (a) Band structure along the high symmetry lines Γ(0, 0, 0) – X(π/*a*, 0, 0) – S(π/*a*, π/*b*, 0) – Y(0, π/*b*, 0) – Γ(0, 0, 0) – Z(0, 0, π/*c*) – U(π/*a*, 0, π/*c*) – R(π/*a*, π/*b*, π/*c*) – T(0, π/*b*, π/*c*) – Z(0, 0, π/*c*). Inset: the first Brillouin zone with the arrows indicating the *k* path for the band structure calculations. (b) Density of states (DOS) projected onto the Hf-4*d* and O-2*p* orbitals. The Fermi energy is aligned to the valence band maximum and is indicated by the horizontal dashed line.

As was discussed above, the $Pca2_1$ group contains a two-fold screw rotation symmetry $S_z$ given by Eq. (3). Applying this transformation twice we obtain

$$S_z^2 : (x, y, z) \rightarrow (x, y, z + c), \quad (6)$$

which is simply translation along the *z*-axis by vector (0, 0, *c*). For spineless system, we have $S_z^2 \psi_\mathbf{k} = e^{ik_z c} \psi_\mathbf{k}$ and hence

$$S_z^2 = e^{ik_z c}. \quad (7)$$

In addition, the system exhibits time-reversal symmetry *T*. Composition of $S_z$ and *T* defines the anti-unitary symmetry operator $\Theta \equiv S_z T$. Since $S_z$ and *T* commute and $T^2 = 1$ for spineless system, we find

$$\Theta^2 = S_z^2 T^2 = e^{ik_z c}. \quad (8)$$

It is evident from Eq. (8) that at the Brillouin zone boundary, $k_z = \pi/c$, the wave function changes sign under this transformation, i.e. $\Theta^2 = -1$. Since Θ is an anti-unitary operator, which commutes with the Hamiltonian and preserves the momentum at $k_z = \pi/c$ plane, the two Bloch states $\psi_\mathbf{k}$ and $\Theta \psi_\mathbf{k}$ are eigenfunctions of the Hamiltonian that are orthogonal and have the same eigenvalue.[51] This implies that all the bands at the $k_z = \pi/c$ plane are doubly degenerate.

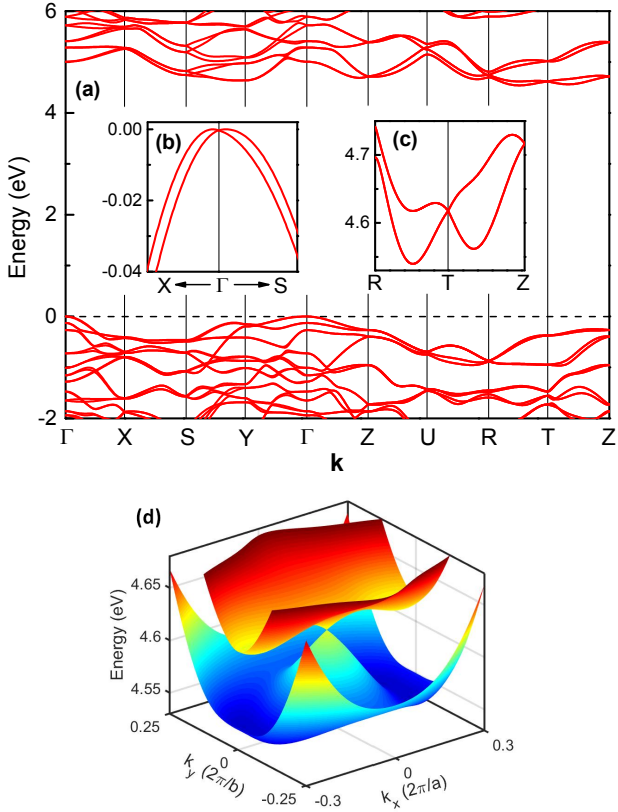

**FIG. 3.** Electronic band structure of HfO$_2$ in presence of SOC. (a) bands along the high symmetry lines (shown in the inset of Fig. 2). Insets: the band structure zoomed in around the Γ point at the valence band maximum (b) and around the T point at the conduction band minimum (c). The dashed line denotes the Fermi energy. (d) Two doubly degenerate bands in the $k_z = \pi/c$ plane around T point (corresponding to $k_x = 0$ and $k_y = 0$ in the plot). The bands cross at the T point forming a 3D Dirac point with fourfold degeneracy.

Fig. 3a shows the calculated relativistic band structure of bulk HfO$_2$. Comparing Figs. 2a and 3a, we see that including spin-orbit coupling leads to sizable splitting of the bands. The splitting is especially pronounced along certain lines in the Brillouin zone, particularly in the $k_z = 0$ and $k_z = \pi/c$ planes. On the other hand, there are special high-symmetry lines and points in the Brillouin zone where the splitting is zero by symmetry. This is, in particular, the case for the Γ–Z line, where $k_x = k_y = 0$ and hence the effective electric field associated with the polar displacements along the z-axis is parallel to the wave vector.

The insets in Fig. 3 show the band structure zoomed in around the VBM and CBM represented by the Γ and T points, respectively. The SOC-induced splitting around the T point (~100 meV) is significantly larger than that around the Γ point (~5 meV) (note different scales in the insets). As we will see in sec. 5, the splitting is Rashba-like around the Γ point, while it is Dresselhaus-like around the T point.

There is another important difference between the bands around the VBM and CBM, resulting from their location at different symmetry points of the Brillouin zone. The bands along the R–T and T–Z symmetry lines preserve their double degeneracy even in the presence of SOC, while the bands along the X–Γ and Γ–S lines are not degenerate (except the Γ point).

The double degeneracy at the Brillouin zone boundary, $k_z = \pi/c$, follows from $\Theta^2 = -1$ which also holds for a spin-half system. In this case, Eq. (7) is replaced by

$$S_z^2 = -e^{ik_z c}, \qquad (9)$$

where the minus sign occurs due to the $S_z^2$ transformation involving a $2\pi$ rotation, which changes sign of the spin-half wave function. In addition, for a non-integer spin system $T^2 = -1$, which in combination with Eq. (9) preserves Eq. (8). Thus, the Θ symmetry provides the double degeneracy for any wave vector **k** in the $k_z = \pi/c$ plane, which is invariant with respect to Θ, also in the spinful system. We note that the double degeneracy is lifted when moving out of the $k_z = \pi/c$ plane except the high symmetry T–Y line.

For the wave vector in the $k_z = \pi/c$ plane around the CBM, there are two doubly degenerate bands crossing at the time reversal invariant T point (Fig. 3c). This crossing and the four-fold degeneracy at this point are protected by the symmetry $M_1$, which is evident from the following consideration.[52] According to Eq. (4), under the $M_1^2$ operation the spatial coordinate transforms as

$$M_1^2 : (x, y, z) \to (x + a, y, z), \qquad (10)$$

whereas the spin component of the wave function changes its sign, which leads to $M_1^2 = -e^{ik_x a}$. Therefore, along the high symmetry line R–T–R invariant under $M_1$ transformation, each band can be labelled by its $M_1$ eigenvalue, $ie^{\frac{i}{2}k_x a}$ or $-ie^{\frac{i}{2}k_x a}$, i.e. $M_1 \psi_\mathbf{k}^\pm = \pm i e^{\frac{i}{2}k_x a} \psi_\mathbf{k}^\pm$. On the other hand, as follows from the commutation relation between Θ and $M_1$, the two degenerate states, $\psi_\mathbf{k}^\pm$ and $\Theta \psi_\mathbf{k}^\pm$, have the same eigenvalue of $M_1$:

$$M_1 \Theta \psi_\mathbf{k}^\pm = -e^{ik_x a} \Theta M_1 \psi_\mathbf{k}^\pm = \pm i e^{\frac{i}{2}k_x a} \Theta \psi_\mathbf{k}^\pm. \qquad (11)$$

Therefore, when two degenerate bands with different $M_1$ eigenvalues cross, the resulting crossing point is protected, leading to four-fold degeneracy. At the same time, since $T$ commutes with $M_1$, the states with different eigenvalues of $M_1$ are connected by the time-reversal symmetry. This is seen from

$$M_1 T \psi_\mathbf{k}^\pm = T[\pm i e^{\frac{i}{2}k_x a} \psi_\mathbf{k}^\pm] = \mp i e^{-\frac{i}{2}k_x a} T \psi_\mathbf{k}^\pm, \qquad (12)$$

which implies that $M_1$ transforms the wave function $T \psi_\mathbf{k}^\pm$ in the same way as $\psi_{-\mathbf{k}}^\mp$, and hence $T \psi_\mathbf{k}^\pm = \psi_{-\mathbf{k}}^\mp$. Therefore, the symmetry protected crossing of the two degenerate bands with different $M_1$ eigenvalues must occur at the time reversal invariant **k**-point, i.e. the T point in our case. Fig. 3d shows the electronic band structure around the T point in the $k_z = \pi/c$ plane. The four-fold degenerate time-reversal invariant T point is in fact a 3D Dirac point.[53]

At the $k_z = 0$ plane $\Theta^2 = 1$, as follows from Eq. (8), so that the wave functions $\psi_\mathbf{k}$ and $\Theta\psi_\mathbf{k}$ are not any longer orthogonal and represent the same state. Therefore, the bands around the VBM (Fig. 3b) are not degenerate, except the Γ point. At the Γ point, the energy level is double degenerate due to **k** = 0 being a time-reversal invariant wave vector, which implies that the **k** = 0 state is a Kramers doublet.

## V. SPIN STRUCTURE

Now we focus on the spin structure of the bands around the Γ point (at the VBM) and the T point (near the CBM). In both cases, a constant energy line crosses bands with four different wave vectors, two being closer to and two being further from the symmetry point. It is convenient to distinguish these bands as 'inner' and 'outer' branches. Figs. 4a and 4b show the calculated spin structure for the two branches around the Γ point in the $k_z = 0$ plane. The out-of-plane spin component, $s_z$, is zero by symmetry, while the in-plane spin components, $s_x$ and $s_y$, display a pronounced chiral spin texture. The chirality changes from counter-clockwise for the inner branch to clockwise for the outer branch. It is seen that both for inner and outer branches the spin is orthogonal to the wave vector **k**, which is typical for the Rashba-type SOC.

The spin structure around the T point in the $k_z = \pi/c$ plane exhibits distinctly different features. As is seen from Figs. 4c and 4d, the angle between **k**-vector and the spin depends on the direction of **k**. The spin is perpendicular to **k** along the $k_x$ and $k_y$ axes but parallel to **k** along the diagonals (the T–U direction in the Brillouin zone). This behavior is typical for the Dresselhaus-type SOC. Furthermore, we see the presence of a sizable out-of-plane spin component, $s_z$, which is indicated by color in Figs. 4c and 4d. All three spin components, i.e. $s_x$, $s_y$, and $s_z$, change sign between the inner and outer branches, reflecting change of the spin direction with respect to the effective SOC field $\mathbf{\Omega}(\mathbf{k})$ at a particular **k** point (Eq. (1)). The $s_z$ also changes sign when crossing the $k_y = 0$ line, as will be explained below using a model Hamiltonian.

As we saw, even in the presence of SOC each of the spin-split bands represents a doubly degenerate state in the $k_z = \pi/c$ plane. Figs. 4e and 4f show the respective spin structures around the T point for the doublet-conjugated states for inner and outer branches. By comparing Fig. 4c to 4e and Fig. 4d to 4f, we see that the doubly degenerate states possess the same in-plane spin components but opposite out-of-plane spin components. This is explained by non-symmorphic symmetry of the crystal combined with time-reversal symmetry, as represented by the $\Theta \equiv S_z T$ operator, which transforms the states within each of the two doubly degenerate bands. The $S_z$ transforms the spin components according to $(s_x, s_y, s_z) \rightarrow (-s_x, -s_y, s_z)$, whereas $T$ transforms **s** to −**s**. Combing the two transformations, we find that within each degenerate band the two spin states have opposite $s_z$ but the same $s_x$ and $s_y$. We note that due to equal population of the Θ-conjugated states in each of the two doubly degenerate bands, the ensemble-average value of the out-of-plane spin component, $s_z$, is zero, whereas the in-plane spin components, $s_x$ and $s_y$, remain finite.

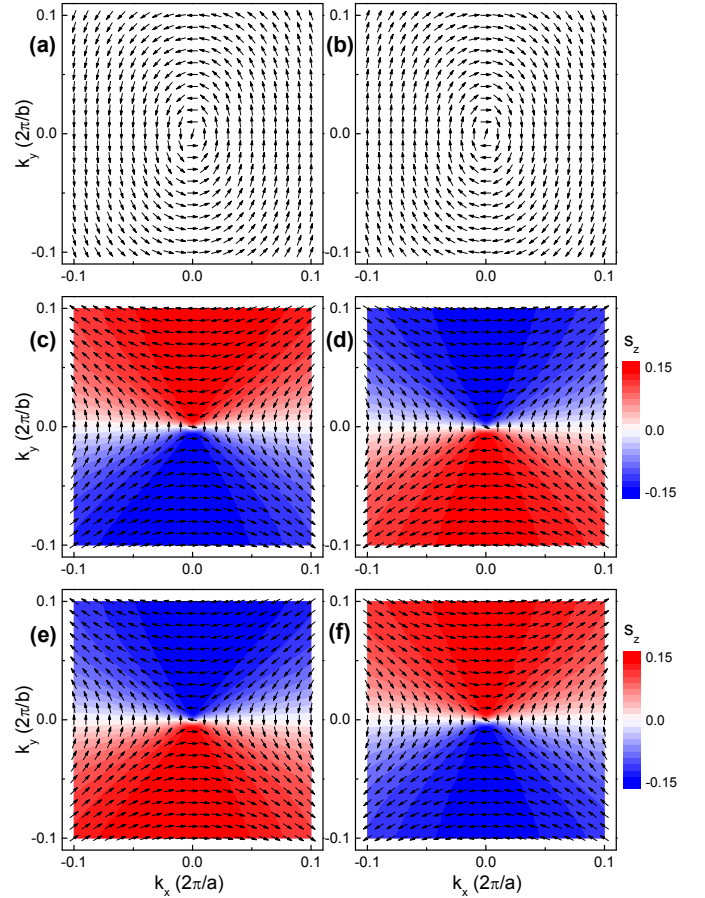

**FIG. 4.** Calculated spin textures in the $k_z = 0$ plane around the Γ point at the top of the valence band (a, b) and in the $k_z = \pi/c$ plane around the T point at the bottom of the conduction band (c-f) for the inner branches (a, c, e) and the outer branches (b, d, f). Panels (c, e) and (d, f) corresponds to the two conjugated states within the double degenerate band. The in-plane spin components, $s_x$ and $s_y$, are shown by the arrows while the out-of-plane spin component, $s_z$, is indicated by color. The reference $k$-point ($k_x = k_y = 0$) corresponds to the Γ point in (a-b) and the T point in (c-f).

We also note that our DFT calculations indicate that around the T point the two Θ-conjugated states in each doubly degenerate band are composed of the *d* orbitals localized on the Hf atoms lying in the two different atomic layers which are separated along the *c* direction (Fig. 1b). Since the $\Theta$-conjugated states have opposite $s_z$ components, this spatial separation creates a local spin-polarization at a given **k**-point. This behavior is reminiscent to that tagged as 'hidden spin-polarization.'[54]

## VI. A MODEL

The spin textures around the $\Gamma$ and T points can be understood in terms of an effective $\mathbf{k} \cdot \mathbf{p}$ Hamiltonian, which can be deduced from symmetry considerations. The wave-vector symmetry group of the $Pca2_1$ space group at the $\Gamma$ point is $C_{2v}$, which has two-fold rotation $C_{2z}$ around the *z* axis as well as two mirror reflections about the *xz*-plane ($M_y$) and *yz*-plane ($M_x$). The corresponding transformations for **k** and $\boldsymbol{\sigma}$ are given in Table 2. To make the SOC Hamiltonian (1) invariant under these transformations in the $k_z = 0$ plane, the effective SOC field must have form of $\boldsymbol{\Omega}(\mathbf{k}) = (\alpha k_y, \beta k_x, 0)$, where $\alpha$ and $\beta$ are some constants. Note that the symmetry forbids having linear in **k** components proportional to $\sigma_z$ and thus the out-of-plane spin component is zero.

Taking into account these considerations, the effective Hamiltonian characterizing the electronic and spin structure in the $k_z = 0$ plane around the $\Gamma$ point can be written as follows:

$$H = E_0 + H_{SO}, \quad (13)$$

where

$$E_0 = \frac{\hbar^2 k_x^2}{2m_x} + \frac{\hbar^2 k_y^2}{2m_y} \quad (14)$$

is the free-electron contribution with $m_x$ ($m_y$) being the electron effective mass along the $k_x$ ($k_y$) direction, and

$$H_{SO} = \alpha k_x \sigma_y + \beta k_y \sigma_x \quad (15)$$

is the SOC coupling term. The latter includes the Rashba and Dresselhaus SOC effects, as can be seen from rewriting Eq.(15) in form $H_{SO} = \lambda_D \left( k_x \sigma_y + k_y \sigma_x \right) + \lambda_R \left( k_x \sigma_y - k_y \sigma_x \right)$, where $\lambda_D = (\alpha + \beta)/2$ and $\lambda_R = (\alpha - \beta)/2$ are the Dresselhaus and Rashba parameters, respectively.

**Table 2.** Transformation rules for wave vector **k**, and spin ($\boldsymbol{\sigma}$) and sublattice ($\boldsymbol{\tau}$) Pauli matrices under the $C_{2v}$ point-group symmetry operations for the $\Gamma$ (0, 0, 0) and T (0, $\pi/b$, $\pi/c$) points in the Brillouin zone of $HfO_2$ (shown in the inset of Fig. 2a). The wave vector **k** is referenced with respect to the high symmetry point ($\Gamma$ and T) where it is assumed to be zero. *K* denotes complex conjugation.

| $\Gamma$ point | | | | T point | | | |
|---|---|---|---|---|---|---|---|
| Symmetry operation | $(k_x, k_y, k_z)$ | $(\sigma_x, \sigma_y, \sigma_z)$ | | Symmetry operation | $(k_x, k_y, k_z)$ | $(\sigma_x, \sigma_y, \sigma_z)$ | $(\tau_x, \tau_y, \tau_z)$ |
| $T = i\sigma_y K$ | $(-k_x, -k_y, -k_z)$ | $(-\sigma_x, -\sigma_y, -\sigma_z)$ | | $T = i\sigma_y \tau_z K$ | $(-k_x, -k_y, -k_z)$ | $(-\sigma_x, -\sigma_y, -\sigma_z)$ | $(-\tau_x, \tau_y, \tau_z)$ |
| $S_z = i\sigma_z$ | $(-k_x, -k_y, k_z)$ | $(-\sigma_x, -\sigma_y, \sigma_z)$ | | $S_z = \sigma_z \tau_x$ | $(-k_x, -k_y, k_z)$ | $(-\sigma_x, -\sigma_y, \sigma_z)$ | $(\tau_x, -\tau_y, -\tau_z)$ |
| $M_1 = i\sigma_y$ | $(k_x, -k_y, k_z)$ | $(-\sigma_x, \sigma_y, -\sigma_z)$ | | $M_1 = i\sigma_y$ | $(k_x, -k_y, k_z)$ | $(-\sigma_x, \sigma_y, -\sigma_z)$ | $(\tau_x, \tau_y, \tau_z)$ |
| $M_2 = i\sigma_x$ | $(-k_x, k_y, k_z)$ | $(\sigma_x, -\sigma_y, -\sigma_z)$ | | $M_2 = \sigma_x \tau_x$ | $(-k_x, k_y, k_z)$ | $(\sigma_x, -\sigma_y, -\sigma_z)$ | $(\tau_x, -\tau_y, -\tau_z)$ |

The band energies of Eq. (13) are given by

$$E_k^\pm = E_0 \pm \sqrt{\alpha^2 k_x^2 + \beta^2 k_y^2}. \quad (16)$$

By fitting the DFT calculated band structure around the $\Gamma$ point, we find for the Rashba and Dresselhaus parameters: $\lambda_R = 0.056$ eV Å and $\lambda_D = 0.007$ eV Å.

For the T point, the situation is different. Here, additional sublattice degrees of freedom need to be included in the consideration to take into account four dispersing bands. This is conventionally described by a set of Pauli matrices $\tau_\alpha$ operating in the sublattice space.[52] The wave-vector symmetry group of the $Pca2_1$ space group at the T point is still $C_{2v}$, but the symmetry operations now include transformations both in the spin and sublattice space. These are given in Table 2. Collecting all the terms which are invariant with respect to the symmetry operations we obtain the effective Hamiltonian as follows:

$$H_{SO} = \alpha k_x \sigma_y + \beta k_y \sigma_x + k_y \sigma_z (\gamma_1 \tau_y + \gamma_2 \tau_z) + \\ + k_z \sigma_y (\delta_1 \tau_y + \delta_2 \tau_z) + \chi k_z \tau_x. \quad (17)$$

Here $\gamma_1$, $\gamma_2$, $\delta_1$, $\delta_2$, and $\chi$ are constants, and the wave vector **k** is referenced with respect to the T point where it is assumed to be zero. This Hamiltonian guarantees double degeneracy of the bands in the $k_z = \pi/c$ plane (corresponding to $k_z = 0$ in Eq. (17)), due to symmetry protection: $(S_z T)^2 = \left(-i\sigma_x \tau_y K\right)^2 = -1$. Moving away from that plane ($k_z \neq 0$ in Eq.(17)) breaks the double degeneracy and splits the bands into four singlets except the $k_x = k_y = 0$ line where the double degeneracy is preserved.

For $k_z = 0$, the Hamiltonian (17) can be easily diagonalized in the sublattice space. The eigenvalues of the $\gamma_1 \tau_y + \gamma_2 \tau_z$ matrix are $\eta \gamma$, where $\gamma = \sqrt{\gamma_1^2 + \gamma_2^2}$ and $\eta = \pm 1$. Thus, the effective SOC Hamiltonian around the T point in the $k_z = \pi/c$ plane can be represented in form

$$H_{SO} = \alpha k_x \sigma_y + \beta k_y \sigma_x + \eta \gamma k_y \sigma_z. \qquad (18)$$

This is equivalent to the representation (1) with the SOC field $\mathbf{\Omega}_\eta(\mathbf{k}) = (\alpha k_y, \beta k_x, \eta \gamma k_y)$, which has opposite sign of the $z$ component for different values of $\eta = \pm 1$. The band energies $E_k$ are degenerate with respect to $\eta$, producing two doublets with energies

$$E_k^\pm = E_0 \pm E_{SO}, \qquad (19)$$

where $E_{SO} = \sqrt{\alpha^2 k_x^2 + (\beta^2 + \gamma^2) k_y^2}$. The normalized spinor wave function $\psi_k$ is given by

$$\psi_k^\pm = \frac{e^{i\mathbf{k}\cdot\mathbf{r}}}{\sqrt{2\pi(\rho_\pm^2 + 1)}} \begin{pmatrix} \dfrac{i\alpha k_x - \beta k_y}{\eta \gamma k_y \mp E_{SO}} \\ 1 \end{pmatrix}, \qquad (20)$$

where $\rho_\pm^2 = \dfrac{\alpha^2 k_x^2 + \beta^2 k_y^2}{(\eta \gamma k_y \mp E_{SO})^2}$.[55] The expectation value of the spin operator is obtained from $\mathbf{s}^\pm = \tfrac{1}{2}\langle \psi_k^\pm | \boldsymbol{\sigma} | \psi_k^\pm \rangle$, resulting in

$$(s_x, s_y, s_z)^\pm = \pm \frac{1}{2 E_{SO}} (\beta k_y, \alpha k_x, \eta \gamma k_y). \qquad (21)$$

As is evident from Eq. (21), the states with different values of $\eta = \pm 1$ have opposite sign of the $z$ component of the spin. Note that the $s_z$ changes sign when crossing the $k_y = 0$ line consistent with the DFT calculations (Figs. 4(c-f)).

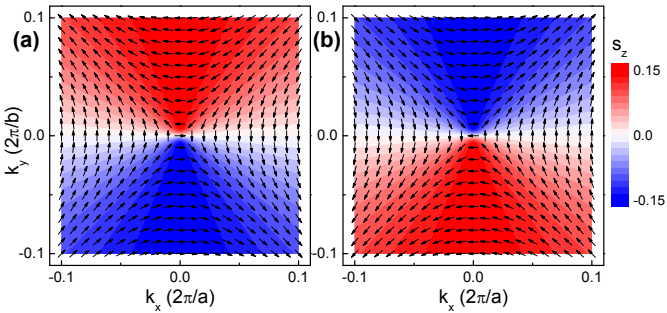

**FIG. 5.** Calcuated spin textures based on the model of Eq. (18) for inner $\psi_k^+$ (a) and for outer $\psi_k^-$ (b) branches. The in-plane spin components, $s_x$ and $s_y$, are shown by the arrows while the out-of-plane spin component $s_z$ is indicated by color.

Using Eq. (19), we fit the electronic band structure around the T point and obtain the following parameters of the model Hamiltonian (18): $\alpha$ = 0.605 eV Å, $\beta$ = 0.550 eV Å, $\gamma$ = 0.168 eV Å. From $\alpha$ and $\beta$ we find the Rashba and Dresselhaus parameters: $\lambda_R$ = 0.028 eV Å and $\lambda_D$ = 0.578 eV Å. It is evident that the Dresselhaus SOC splitting dominates the Rashba SOC splitting around the T point. The large Dresselhaus constant explains the spin texture found from our DFT calculation and shown in Figs. 4 (c-f). This behavior is nicely reproduced by the $\mathbf{k}\cdot\mathbf{p}$ model with the $\lambda_R$ and $\lambda_D$ extracted from the DFT calculation. The respective spin textures for $\eta$ =1 are shown in Fig. 5, indicating the Dresselhaus-type feature in spin configuration. The conjugated doublet state with $\eta = -1$ has the same in plane spin component but opposite sign of $s_z$ (not shown).

## VII. OUTLOOK

The Dresselhaus effect is generally found in the bulk materials with spatial inversion symmetry broken, such as zinc-blende semiconductors. For comparison, we summarize in Table 3 the Dresselhaus parameters $\lambda_D$ obtained from DFT calculations for a few selected bulk systems. We see that the predicted value of $\lambda_D$ for bulk $HfO_2$ is significantly larger than the values known for non-organic bulk semiconductors and oxides. The large $\lambda_D$ reported for organic $(FA)SnI_3$ assumes a specific non-centrosymmetric crystal structure of this material, which may be difficult to realize in practice due to disorder in organic cation dipole moments.

**Table 3.** Calculated Dresselhaus parameters for selected bulk materials.

| Material | $\alpha_D$ (eV Å) | Reference |
|---|---|---|
| $HfO_2$ ($Pca2_1$) | 0.578 | This work |
| AlAs | 0.011 | [21] |
| GaP | 0.072 | [21] |
| $BiAlO_3$ ($R3c$) | 0.041 | [31] |
| $(FA)SnI_3$ | 1.190 | [26] |

$HfO_2$ is a wide band gap material which can be used as an insulating barrier in tunnel junctions. The ferroelectric phase of this material is especially interesting due to the tunneling electroresistance effect[56,57] known as an important functional property of ferroelectric and multiferroic tunnel junctions.[58] Due to $HfO_2$ being well compatible with the existing semiconductor technologies, it can potentially be employed to develop FTJ-based memories.

The presence of the large SOC coupling effects predicted in this work opens additional interesting possibilities for using this material. In ferroelectric $HfO_2$, a full reversal of the spin texture is expected in response to reversal of its ferroelectric polarization **P**, similar to what was originally proposed for GeTe.[23] This behavior follows from the fact that reversal of **P**, i.e. change of **P** to –**P**, is equivalent to the space inversion operation which changes the wave vector from **k** to –**k** but preserves the spin **σ**. Applying the time-reversal symmetry operation to this state with reversed polarization, we bring –**k** back **k** but flip the spin, changing it from **σ** to –**σ**. Thus, the reversed-polarization state is identical to the original state with the same **k** but reversed spin **σ**.

A possible implication of this effect may be found in tunnel junctions. [59] The Rashba SOC at the interface in a magnetic tunnel junction has been predicted to produce a tunneling spin Hall effect and tunneling anomalous Hall effect (AHE).[24] This prediction was extended to the presence of the bulk Dresselhaus contribution in a tunnel junction with a single ferromagnetic electrode.[25] In particular, it was found that magnitude of the tunneling AHE scales linear with the Dresselhaus parameter. The large value of $\lambda_D$ in HfO$_2$ makes this material a favorable candidate for observing this effect experimentally. The presence of ferroelectric polarization causes the AHE to be reversible, because its sign changes with the sign of $\lambda_D$ and hence **P**.

## VIII. SUMMARY

In summary, we have investigated the Rashba and Dresselhaus effects in the bulk ferroelectric oxide HfO$_2$ using first-principles calculations and a $\mathbf{k} \cdot \mathbf{p}$ Hamiltonian model. We focused on the orthorhombic $Pca2_1$ structural phase of HfO$_2$ which was previously predicted theoretically[41] and confirmed experimentally.[42] We found that the calculated structural parameters and ferroelectric polarization are consistent with those reported previously. Results of our calculations showed that ferroelectric HfO$_2$ is an indirect-band-gap insulator with the VBM located at the Γ point and the CBM located near the high symmetry T point. The band energies are doubly degenerate in the $k_z = \pi/c$ plane of the Brillouin zone, which stems from the non-symmorphic space group of the crystal. We found that the time-reversal invariant T point supports band crossings and the four-fold degeneracy of the electronic states which is protected by the crystal symmetry. The calculated spin textures reveal that the Rashba-type SOC dominates around the VBM, whereas the Dresselhaus-type SOC dominates around the CBM. The spin splitting induced by SOC as well as the spin textures are explained by the $\mathbf{k} \cdot \mathbf{p}$ Hamiltonian deduced from symmetry arguments. Importantly, a very large Dresselhaus parameter of 0.578 eV Å is predicted for the orthorhombic HfO$_2$, which is at least an order of magnitude larger than that known for conventional semiconductors and oxides. The spin textures are fully reversible with polarization switching, which enables the control of spin-dependent properties by electric fields. The large Dresselhaus parameter and the reversible spin structure may have interesting implications for ferroelectric tunnel junctions based on HfO$_2$, where sizable spin and anomalous Hall effects are expected, reversible with ferroelectric polarization. Overall, our results provide the fundamental understanding of the Rashba and Dresselhaus effects in ferroelectric HfO$_2$, revealing new functionalities of this material which could be explored experimentally.

## ACKNOWLEDGMENTS

This work was supported by the National Science Foundation (NSF) through Nebraska Materials Research Science and Engineering Center (MRSEC) (NSF Grant No. DMR-1420645). Computations were performed at the University of Nebraska Holland Computing Center. The atomic structures in Fig. 1 were produced using VESTA software.[60]

* E-mail: ltao2@unl.edu  
† E-mail: tsymbal@unl.edu